# Al-Cu-Fe quasicrystals as anode of lithium ion battery

Xiao Lan[a], Haijuan Wang[a], Zhanhao Sun[a], Xunyong Jiang[a,*]

e-mail: jiangxunyong@tjut.edu.cn

[a] School of Materials Science and Engineering, Tianjin University of Technology ,Tianjin, China

**abstract:** In this paper, Al-Cu-Fe quasicrystal alloy was used as anode material of lithium-ion batteries. The first specific discharge capacity of quasicrystal was 204mAh/g. Cyclic voltammetry showed that oxidation peak of Al-Cu-Fe quasicrystal was at about 1.4V. The reduction peak was at 0.3V. Al-Cu-Fe quasicrystal have higher Li-ion diffusion impedance and Warburg in the first cycle. X-ray diffraction analysis demonstrate that Li atom enter into quasicrystal structure can not fully leave quasicrytal during first charge/discharge cycle which induce the irreversible capacity.

## 1. Introduction

Lithium-ion batteries (LIBs) are considered as the state-of-the art energy storage systems due to their high gravimetric and volumetric energy densities. However, the energy density of the current LIBs can not satisfy the ever-growing demands for the electric vehicles (EVs), portable electronics and large-scale renewable energy storage[1]. Compared with commercial graphite anode(372mAh/g) with intercalation reaction mechanism, higher lithium storage capacity could be achieved in the metallic or semimetallic elements (Al, Si, Ge, Sn, Sb, etc.) or in the $M_aX_b$-type compounds (M being a transition metal and X being F, O, P, N, S …)by alloying reaction or conversion reaction with lithium, respectively[2]. Silicon based anodes have the advantages of high capacity and low charge potential. However, the large volume expansion(300％) and the continuous formation of the SEI film during the delithiation process seriously hinder the practical application of the silicon-based anode. The application of the other material is limited by low cycle property , low coulomb efficiency. The search of new ecological metallic anode is one of the important technical route to develop high energy LIB[3].

Quasicrystals(QCs) is a structure that is ordered but not periodic which present 5-fold, or even 8-fold or 10-fold symmetries[4]. The special atomic arrangement of QCs determine its special property. QCs have been used as electrochemical hydrogen storage materials[5]. Al-Cu-Fe QCs, such as $Al_{63}Cu_{25}Fe_{12}$, contain large portion of Aluminum which has high lithium storage property capacity. AlCuFe QCs may storage lithium reversibly. According to our knowledge, there has no study on the lithium storage of Al-Cu-Fe QCs. In this paper, $Al_{63}Cu_{25}Fe_{12}$ was prepared by general melt casting method and high-energy ball milling. The lithium storage property of Al-Cu-Fe QCs was studied. The result show that Al-Cu-Fe QCs has certain lithium storage capacity.

## 2. Materials and methods

$Al_{63}Cu_{25}Fe_{12}$ ingot were prepared by art melting with Fe (chip 99.9%), Cu (shot 99.9%) and Al (shot 99.9%). The ingot was annealed at 800°C for 10h under an argon atmosphere. Then it is put into ball mill to crush into powder. The final particle size were less than 0.0374mm. The electrodes for electrochemical test were prepared as follow: the alloy powder and PVDF were mixed with the mass ratios of 8:1. The mixture with addition of a suitable amount of N-methyl-2-pyrrolidone (NMP) were stirred with a magnetic whisk for 1h, and pasted onto the copper foils. The materials were punched into circular electrode slices (15mm in diameter) after they were dry for 4h at 25°C. Then the electrode slices were placed on a wild mouthed petri dish to be heated for 10h at 120°C in the vacuum drying oven. The average weight of the electrodes was approximately 18 mg. The mental Lithium was used as anode in the cells. And a porous polymeric membrane (Celgard 2400) was used as diaphragm. 1M LiBF4 dissolved in EC, DEC and EMC mixture of equal volumes was used as electrolyte. The cells were assembled in an argon-filled glove box with the concentrations of moisture and oxygen below 0.1 ppm.

The samples were characterized using Rigaku D/Max-2500V X-ray diffraction (XRD) in the range of 10°-90°. Charge-discharge cycles test of the electrodes were carried out on the NEWARE tester at 25°C. IM6 electrochemical workstation was utilized to measure electrochemical impedance spectra (EIS) in the frequency range of 110kHz to 0.01Hz with AC amplitude 5mV. Cyclic voltammetry measurements were tested at a scan rate of 0.1 mV/s in the voltage range of 0-3V (versus Li/Li$^+$) and current range of -40 to 40mA.

## 2. Results and discussion

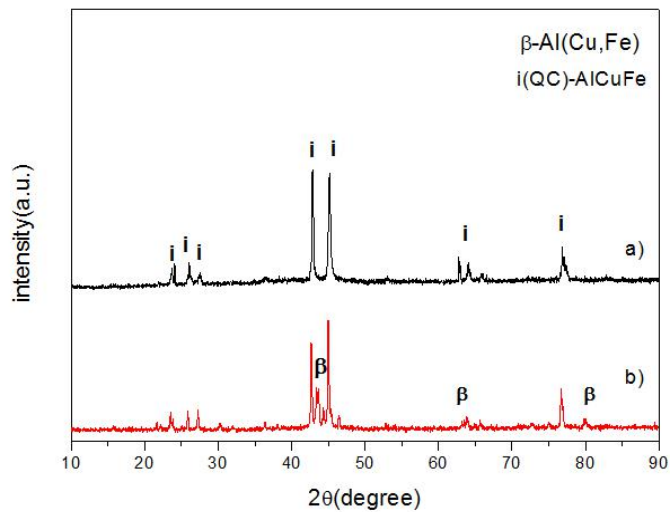

Figure 1 XRD patterns of Al-Cu-Fe alloy powders (a: after heat treatment and ball mill, b: as cast )

Fig.1 is the XRD patterns of Al-Cu-Fe alloy powders(a: after heat treatment and ball mill, b:as cast). For as cast alloy ,there exist two diffraction peak of i phase between 42° and 46° which is diffraction peaks of Al-Cu-Fe QCs [6]. The crystalline β-Al(Cu,Fe) phase also exist in as cast alloy. After heat treatment and high-energy ball milling, the XRD peaks of QC phases still exist. But the peaks of β-Al(Cu,Fe) phase disappear. The XRD results show that pure Al-Cu-Fe QCs can be obtained by heat treatment and subsequent high energy ball milling. The following electrochemical test was performed with pure Al-Cu-Fe QCs prepared by process of Fig.1a.

To eliminate its influence on specific capacity, acetylene black was not added into the anodes as the conducting material. Fig. 2a display the charge-discharge voltage profiles of the Al-Cu-Fe QCs from the 1nd to 50th cycles at a current density of 100mA/g. In the first discharge process, there was a relatively slow discharge platform between 1.0V and 0.5V, corresponding to the formation of SEI film. The tilt table at 0.25V corresponded to the process of lithium ion embedding into QCs material. After the 50th cycle, the platform almost disappeared. The first specific discharge and charge capacities were 204mAh/g and 60mAh/g. There exist the irreversible process during the first discharge process. For the subsequent cycle, the specific discharge and charge capacities were almost same (Fig. 2b) .After 50th cycle, discharge capacities is about 65mAh/g.

CV curves reveal the lithiation/delithiation process of the Al-Cu-Fe QCs architecture. The reduction and oxidation peaks are shown clearly in the CV cure (Fig. 2c). In the first scan, the broad reduction peak between 0.6V and 1.3V is related to the formation of SEI layer. The peaks at 0.2V and 1.4V are related to the intercalation/deintercalation of Li in the active materials to form $AlLi_x$, corresponding to reversible process[7-9]. After the third scan, the broad reduction peak between 0.6V and 1.3V nearly disappeared, indicating a stable SEI film forming on anode surface. There is consistent with the Charge-discharge result, which an QCs has an oblique platform in the first cycle.

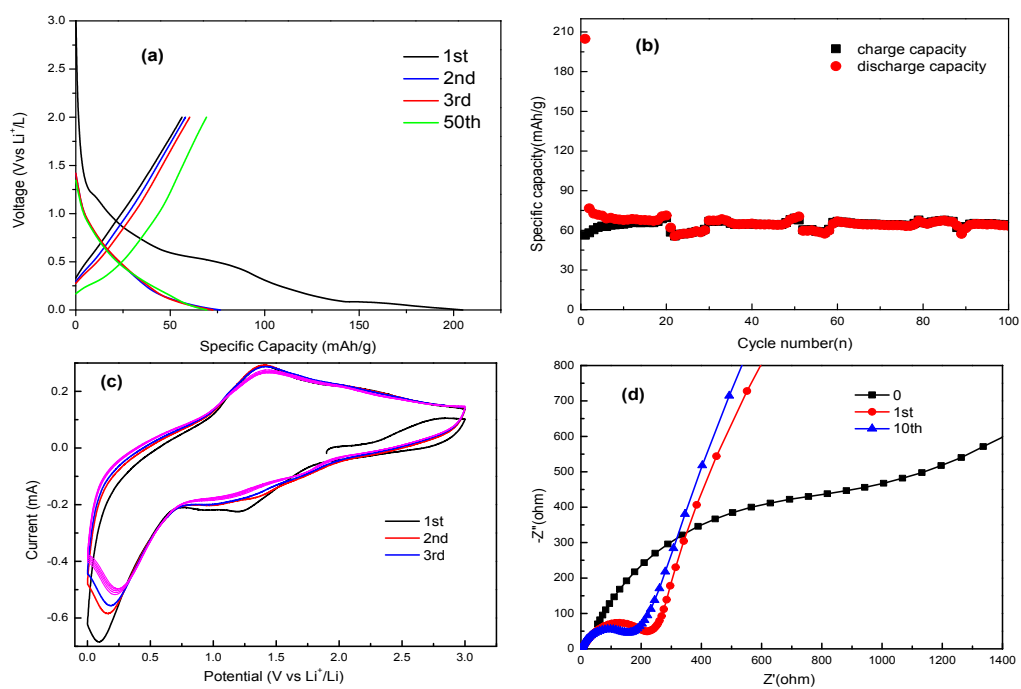

Figure 2 Electrochemical test of Al-Cu-Fe QCs(a) charge-discharge curves. (b) cycling performance at 100mA/g (c) CV curves. (d) Nyquist plots of QCs at an open-circuit potential with different cycle.

EIS results are shown in Fig. 2d. The semicircle in the high-frequency region is ascribed to the formation of SEI film and the process of lithium ion passing through the SEI film, and the oblique line is consistent with the process of lithium ion diffusion in the anode. Initial materials(mark as 0) show the higher charge transfer resistance and higher Warburg impedance than after the first and tenth cycle. With charge-discharge cycling, the semicircle and slop of oblique line becoming smaller, which mean that li-ion transmission channels are opened. The Electrode material structure are more suitable for intercalation/deintercalation of Li-ion. There are no significant difference between the first and tenth cycle in the semicircle, indicating the impedance of Li-ion diffusion in the materials has no significant change. The behaviors of Li in QCs material anode is mixed control of charge transfer and diffusion steps. EIS results show that addition of acetylene black or modification of the material may improve the electrochemical performance of anode.

In order to observe the structure change of QCs during charge-discharge cycle, The cell under different charge state were disassembled in globe box. The prepared slurry was evenly coated on a nickel foam for test battery. The electrode sheets were removed and wrapped with polyethylene film to take the XRD measurement. XRD patterns of the Al-Cu-Fe alloys in the first cycle were shown in the Fig3. The upper right corner of Fig.3 is the charge-discharge curve of AlCuFe QCs. Five samples subjected to XRD measurement in different charge and discharge states are marked in Fig. 3. In order to facilitate the observation of the change of the alloy phase, we have amplified the diffraction peaks of 44° to 45° in the upper left corner of Fig.3 The highly interfering nickel peaks were eliminated. The curve e in Fig.3 is the diffraction peak of the pure protective film, and d is the diffraction peak of the polyethylene film wrapping the QCs alloy. It is obvious that the polyethylene film does not affect the diffraction peak of alloy electrode.

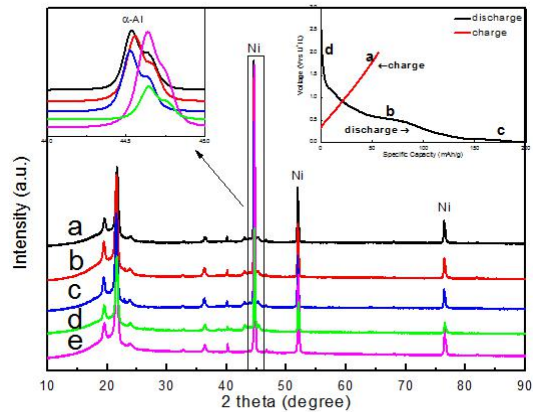

Fig. 3 XRD patterns of AlCuFe QCs in different charge-discharge state and charge-discharge curve in first cycle (upper right corner) (a: charge over, b: discharge to 0.5V, c: discharge over, d: Polyethylene film with alloys, e: Polyethylene film)

The whole charge-discharge cycle is 2V-0V-2V. The half cycle of lithium insertion process is from 2V to 0V which is discharge process. The intermediate point of 0.5V is the middle state of lithium insertion process. It can be seen that(b,c in Fig.3) the peak of the QC phase deviates slightly from the initial position during discharge process. The peaks moved significantly to the left and widened compared with alloy at initial state(d in Fig.3). When discharge process finish, the voltage of anode is 0V. The overall peak shifts toward a small angular direction which mean that the interplanar spacing increases (c in Fig.3). The broadening of the diffraction peaks means decrease in crystalline size and concentration of material stress. However, the intermetallic compound was not shown in the diffraction pattern. This indicate that only the solid solution was present in the alloy when the lithium ions are embedded in the aluminum matrix of the ternary AlCuFe alloy. This induce the small angle movement of the α-Al peak in the XRD pattern. The half cycle is lithium deinsertion process from 0V to 2V which is charge process. When charge completed, the voltage of electrode go back to 2V. The position of the diffraction peak cannot be restored to the original position (a in Fig.3). This mean that part of the lithium ions embedded in the QCs cannot be removed during the deinsertion process. Irreversible capacity loss occurs. The peak shift is small throughout the process. This imply that only the active substances in the surface layer may participate in the reaction.

## 4. Conclusions

In this paper, Al-Cu-Fe QCs was synthesised by melting and subsequent heat treatment. The electrochemical performance of AlCuFe QCs as anodes of LIBs was tested.

1) $Al_{63}Cu_{25}Fe_{12}$ QCs can storage lithium reversibly. There exist the irreversible process during the first discharge process.The first specific discharge capacities were 204mAh/g , dropping to 65mAh/g after 50 cycles.

2) Cyclic voltammogram showed that oxidation peak of Al-Cu-Fe QCs was about 1.4V, and the reduction peak at 0.3V. EIS measurements revealed that the material showed higher Li-ion diffusion impedance and Warburg in the first cycle.

3) During discharging process, lithium ions get into QCs to form solid solution. During charging process part of lithium atom can not leave the QCs which cause irreversible capacity.